\documentclass[aps,a4paper,10pt,twocolumn,nofootinbib]{revtex4} 
\usepackage{amsfonts}
\usepackage{amsmath}
\usepackage{mathrsfs}
\usepackage[mathscr]{euscript}
\usepackage[dvips]{graphicx}
\usepackage{fancyhdr}
\usepackage{colordvi}
\usepackage{url}
\usepackage[hypertex=true]{hyperref}

\def\overstrike#1#2{{\setbox0\hbox{$#2$}\hbox to \wd0{\hss
    $#1$\hss}\kern-\wd0\box0}}

\newcommand{\RealPart}{\mathbb{R}\textrm{e}}
\newcommand{\ImagPart}{\mathbb{I}\textrm{m}}
\newcommand{\ArgPart}{\mathbb{A}\textrm{rg}}

\begin{document}
\title{The refractive index and wave vector in passive or active media}
\author{Paul Kinsler}
\email{Dr.Paul.Kinsler@physics.org}
\affiliation{
  Blackett Laboratory, Imperial College London,
  Prince Consort Road,
  London SW7 2AZ, 
  United Kingdom.
}

\begin{abstract}

Materials that exhibit loss or gain
 have a complex valued refractive index $n$.
Nevertheless, 
 when considering the propagation of optical pulses, 
 using a complex $n$ is generally inconvenient --
 hence the standard choice of 
 real-valued refractive index, 
 i.e.  $n_s = \RealPart ( \sqrt{n^2} )$.
However,
 an analysis of pulse propagation based on the second order wave equation
 shows that use of $n_s$ results in a wave vector
 \emph{different} to that actually exhibited by the propagating pulse.
In contrast, 
 an alternative definition $n_c = \sqrt{ \RealPart ( n^2 ) }$, 
 always correctly provides the wave vector of the pulse.
Although for small loss the difference between the two is negligible, 
 in other cases it is significant; 
 it follows that phase and group velocities are also altered.
This result has implications for the description of 
 pulse propagation in near resonant situations, 
 such as those typical of metamaterials with negative 
 (or otherwise exotic) refractive indices.

\end{abstract}

\lhead{\includegraphics[height=5mm,angle=0]{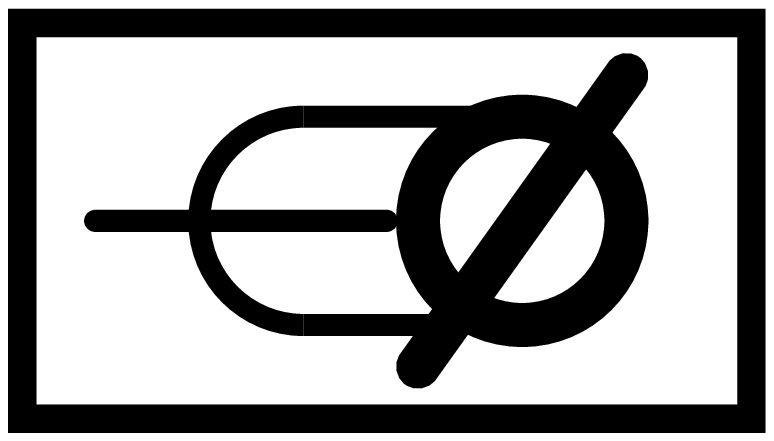}~~VPVG}
\chead{The refractive index and wave vector ...}
\rhead{
Dr.Paul.Kinsler@physics.org\\
http://www.kinsler.org/physics/
}

\date{\today}
\maketitle
\thispagestyle{fancy}

%
\section{Introduction}\label{S-intro}

Recent work in metamaterials
 and negative refractive index media\footnote{Also
  commonly called negative phase velocity (NPV) media} 
 \cite{FocusIssue-2003oe-nrm,Smith-K-2000prl,Shelby-SS-2001s,
  Yen-PFVSPBZ-2004sci,
  Linden-EWZKS-2004sci,Pendry-SS-2006sci,Dolling-EWSL-2006sci,
 Leonhardt-2006sci}
 has focused attention on propagation in media
 with exotic values of permittivity $\epsilon$ and permeability $\mu$, 
 as well as those with significant loss or gain, 
 where $\epsilon$ and $\mu$ are complex valued.
These material properties (i.e. $\epsilon, \mu$) 
 impact directly on the refractive index, 
 and hence on the wave vector $\beta$
 and phase and group velocities \cite{Brillouin-1960-WPGV,Biot-1957pr}.

When considering analytical solutions of the wave equation, 
 it is often convenient to allow the propagation wave vector $\beta$
 and refractive index $n$ to be complex valued,
 based on the definition $n^2 = c^2 \epsilon \mu$, 
 so that $\beta = ( \omega^2 n^2 / c^2 )^{1/2}$.
However, 
 although this leads to many useful results, 
 the approach also has some serious drawbacks.
For example, 
 the sign of the imaginary part of $\beta$,
 which determines whether the wave experiences gain or loss, 
 needs to be specified according to the chosen direction of propagation.  
Worse,
 in the envelope and carrier description of pulse propagation, 
 which is common in nonlinear optics 
 (e.g. see \cite{Agrawal-NFO}), 
 the presence of a complex wave vector in the carrier
 function is very inconvenient, 
 since
 it requires the nonlinear coefficients to be adjusted
 to compensate for the distance propagated.
In addition, 
 determining other parameters such as the group velocity
 under these circumstances is also a non-trivial task 
 (see e.g. \cite{Censor-1977jpa}).  
For these and other reasons, 
 it is often preferable to define a real-valued wave vector $k$ and
 to treat the imaginary component separately.

The standard approach is to simply define $k$
 as the real part of $\beta$, 
 i.e. 
 $k = (\omega/c) \RealPart(\sqrt{n^2}) = \omega n_s/c$.
However, 
 an alternative definition based on 
 $k^2 = (\omega/c)^2 \RealPart \left(n^2\right) = \omega^2 n_c^2/c^2$
 has been used with advantage in studies of causality-based constraints 
 for negative refraction \cite{Stockman-2007prl,Kinsler-M-2008prl}, 
 although neither paper remarked on the non-standard definition.  
In that context, 
 this alternative definition is \emph{required}
 because it keeps the real and imaginary parts of $n^2$ separate,
 and so ensures the Kramers-Kronig relations \cite{LandauLifshitz} 
 continue to hold, 
 linking the two parts and enforcing causality.  
In contrast, 
 the standard complex $n$ is not required to be causal, 
 although it is so in the case of passive (lossy) media
 (see e.g. \cite{Skaar-2006pre,Skaar-2006ol}).

In the present paper, 
 the two definitions will be compared using the
 predictions of the second-order wave equation as the benchmark.  
It is shown that for field propagation
 in media with loss (``passive'') or gain (``active''), 
 where the use of a complex wave vector is particularly problematic, 
 the alternative definition has the clear advantage
 that it exactly matches the spatial oscillations of the field.
In contrast, 
 the standard definition gives an imperfect match, 
 and the description only recovers the true propagation
 due to the presence (and inconvenience) of additional correction terms.
Note that the alternative definition (for $n_c$) is not in any sense 
 equivalent to one based on an effective refractive index, 
 such as might occur in (e.g.) waveguides:
 it is an alternative choice of definition for the bulk refractive index.

Because I focus on the propagation of waves,
 in section \ref{S-waveequation} I present a short description
 of the second order wave equation.
Then, 
 in section \ref{S-definitions},
 I give some definitions
 required for the handling of both
  the standard case (section \ref{S-standard})
 and the new alternative definition (section \ref{S-causal}).
After discussion of the similarities and difference
 between the definitions in section \ref{S-comparisons}, 
 I end by presenting my conclusions in section \ref{S-conclusion}.

%
\section{The second order wave equation}\label{S-waveequation}

The second order wave equation is commonly used in optics
 (at least as a starting point) in descriptions of propagation,
 and results from the substitution of the $\nabla \times \vec{H}$
 Maxwell's equation into the $\nabla \times \vec{E}$ one
 in the source-free case
 (see e.g. \cite{Agrawal-NFO}).
In homogeneous media, 
 with $\nabla^2 = \partial_x^2 + \partial_y^2 + \partial_z^2$
 and $\partial_a \equiv \partial / \partial a$, 
 the frequency space wave equation is
~
\begin{align}
  \nabla^2 \vec{E} 
 +
  \beta^2 \vec{E}
&=
  0
.
\label{eqn-2owe}
\end{align}

Here $\beta^2 = \epsilon \mu \omega^2$
 is the square of a complex propagation wave vector, 
 since both $\epsilon$ and $\mu$ can be complex.
We can relate it to a complex refractive index squared quantity
 with
~
\begin{align}
  \beta^2
&=
  n^2 \frac{\omega^2}{c^2}
.
\label{eqn-2owe-beta2}
\end{align}
When considering the propagation of fields, 
 it is convenient to 
 split $\beta^2$ up into two parts 
 (e.g. its real and imaginary parts).
Here I write $\beta^2 = k^2 + \imath \gamma^2$, 
 so that eqn. (\ref{eqn-2owe}) becomes
~
\begin{align}
  \nabla^2 \vec{E} 
 +
  k^2 \vec{E}
 +
  \imath
  \gamma^2 \vec{E}
&=
  0
.
\label{eqn-2owe-split}
\end{align}
When considering this wave equation, 
 we will usually want the first two terms
 to give plane-wave solutions, 
 with the rest component containing loss
 and nonlinearity\footnote{We can even incorporate diffraction
 in the rest by including the transverse
 parts of $\nabla^2$; 
 see \cite{Kinsler-2008-fchhg}.}.
This is an important step, 
 since although we might solve linear problems 
 using a complex valued $n$, 
 realistic situations are not so easily handled.

The first two terms in eqn. (\ref{eqn-2owe-split}), 
 taken in isolation, 
 have plane-wave solutions if $k$ is real-valued; 
 I call this the ``underlying propagation''.
The third term in eqn. (\ref{eqn-2owe-split})
 is the ``residual'' component,
 which controls the discrepancy between the true propagation
 and the underlying propagation.
Although in the case of small loss or gain
 the residual component will be only a weak perturbation, 
 the theory presented here is valid for \emph{any} strength.

As an aside, 
 if we specialize to the case of fields propagating along the $z$ direction, 
 using the carrier and envelope models of pulse propagation
 \cite{Agrawal-NFO,Brabec-K-1997prl,
       Kinsler-N-2003pra,Genty-KKD-2007oe,Kinsler-2007-envel},
 we would write 
 $E(z,t) = A(z,t) \exp [\imath ( \omega t - k z)] + \textrm{c.c.}$
 to accommodate the rapidly oscillating behaviour
 of the carrier frequency:
 this carrier represents the underlying propagation for a specific frequency.
This then leaves
 only the (usually) slowly varying envelope $A(z,t)$, 
 which would be affected only by the residual component.

Returning to the wave equation of eqn. (\ref{eqn-2owe-split}),
 and taking propagation along the $z$-axis, 
 we can now factorize it using Greens functions
 \cite{Ferrando-ZCBM-2005pre,Kinsler-2007-envel,Genty-KKD-2007oe},
 to give two first-order equations that are coupled 
 only by the residual component.
At the same time we can split the field
 into forward ($E_+$) and backward ($E_-$) parts
 (i.e. set $E = E_+ + E_-$), 
 to give a pair of coupled, 
 counter propagating, 
 first order differential equations.
These are
~
\begin{align}
  \partial_z E_{\pm}
&=
 \pm
  \imath k E_{\pm}
 \mp
  \frac{\gamma^2}
       {2 k}
  \left(
    E_+ + E_-
  \right)
.
\label{eqn-waveequation-factored}
\end{align}
Here the underlying propagation is, 
 as desired, 
 plane-wave like, 
 since the first RHS term just adds an $\imath k z$ behaviour
 onto the frequency dependent $\imath \omega t$.
The propagation is then modified by the second RHS term, 
 i.e. the $\gamma^2$-dependent residual component.
A feature of this approach is that we see that 
 \emph{any} contribution 
 (whether linear or not) 
 that is included in the residual component
 will couple the forward and backward fields together
 (see 
 \cite{Kinsler-2007-envel,Genty-KKD-2007oe} for more discussion). 
Since such terms are scaled by $k$ in eqn. (\ref{eqn-waveequation-factored}), 
 they change (but in a simple way) under  
 my alternative form for the refractive index.

Here I consider only the one dimensional linear case, 
 where $\beta^2$ is independent of the field.
This covers the cases of both loss and/or gain
 (i.e. in passive and/or active media); 
 however for simplicity I will often only refer to loss; 
 nevertheless the case of gain is always allowed for
 (since gain can be seen as ``negative loss'').

If we take the propagation to be of the form
 $E_+ = E_0 \exp \left[ \imath ( \omega t - k' z ) \right]$, 
 with $E_-=0$, 
 then eqn. (\ref{eqn-waveequation-factored}) gives us 
~
\begin{align}
 -
  \imath k'
&=
 -
  \imath k 
 +
  \frac{\gamma^2}
       {2 k}
,
\end{align}
so that $\gamma^2 <0$ corresponds to loss
 for a forward propagating wave.
Further, 
 if we consider instead the oppositely propagating wave, 
 eqn.(\ref{eqn-waveequation-factored}) automatically ensures 
 the necessary change of sign to ensure a loss stays loss, 
 and a gain stays a gain.
In contrast, 
 when using a complex-valued $n$, 
 care must be taken to ensure the correct sign
 (see e.g. \cite{Nistad-S-2008pre}).

%
\section{Definitions}\label{S-definitions}

We have that $\beta^2$ and $n^2$ are (in general) complex valued, 
 and $\omega$ and $c$ are strictly real valued.
Thus when choosing the propagation wave vector
 we need to decide what to do about the imaginary parts.
Our choice then affects the performance,
 utility,
 and convenience
 of the refractive index,
 phase velocity, 
 and group velocity.

I now define some useful intermediate quantities to express
 the refractive index conveniently; 
 I introduce $n_0^2 = | n^2 |$
 and the angle $\phi = \ArgPart (n^2)$ so that 
~
\begin{align}
  n^2
&=
  n_0^2 e^{\imath \phi}
,
\\
  n
&=
  n_0 e^{\imath \phi/2}
.
\end{align}
Whether or not specific values of $\phi$ correspond
 to a negative refractive index or negative phase velocity can be determined
 from the criteria for $\epsilon$ and $\mu$
 given in \cite{Depine-L-2004motl}\footnote{Note that  
 the $\phi$ used here corresponds to $\phi_+$
 in the summary in \cite{Kinsler-M-2008motl}}.
I also define a reference wave vector $k_n$ such that 
~
\begin{align}
  k_n^2
&=
  \frac{\omega^2}{c^2}
  n_0^2
.
\end{align}

The standard form for a real valued refractive index is 
~
\begin{align}
  n_s
=
  \RealPart \left( \sqrt{\left(n^2\right)} \right)
\qquad
&=
  n_0 \cos \frac{\phi}{2}
.
\label{eqn-defs-standard-n}
\end{align}
I have already noted that many treatments leave $n$
 as a complex valued quantity, 
 leading to a complex wave vector $k$; 
 and that while useful in many circumstances, 
 in the context of pulse propagation it brings 
 some significant disavantages.

An alternative definition for the refractive index is
~
\begin{align}
  n_c
&=
  \sqrt{
    \RealPart
      \left(
        n^2 
      \right)
  }
\qquad
=
  n_0 \sqrt{\cos \phi}
,
\label{eqn-defs-causal-n}
\end{align}
where $n_c^2$ satisfies the Kramers-Kronig relations \cite{LandauLifshitz}
 in partnership with the imaginary part $\ImagPart \left(n^2\right)$; 
 this definition has already been used in the literature
 (e.g. see the recent \cite{Stockman-2007prl,Kinsler-M-2008prl}).

%
\section{The standard form}\label{S-standard}

The standard form for the wave vector
 based on the standard form of refractive index
 (see eqn. (\ref{eqn-defs-standard-n})),
~
\begin{align}
  k_s^2
&=
  \frac{\omega^2}{c^2}
  \left[
    \RealPart
      \left(
        \sqrt{n^2}
      \right)
  \right]^2
\qquad
=
  k_n^2 \cos^2 \frac{\phi}{2}
\\
  k_s
&=
  k_n \cos \frac{\phi}{2}
.
\label{eqn-standard-k}
\end{align}
Thus $k_s$ is always real-valued, 
 and can be negative in some circumstances.
The phase velocity is then the usual $v_p = c / n_s$, 
 and the (inverse) group velocity simply $v_g^{-1}=\frac{dk}{d\omega}$.

Let us now consider how this standard form of $k_s^2$
 looks when substituted into the second order wave equation.
To do this let us express $\beta^2$ in terms of $k_s^2$ and $k_n^2$, 
~
\begin{align}
  \beta^2
&=
  k_s^2
 +
  \imath
  k_n^2 
  \gamma_s^2
.
\label{eqn-standard-beta2gamma2}
\end{align}
 with the residual behaviour described by 
~
\begin{align}
  \imath
  \gamma_s^2
&=
  \imath
  \left[
    \sin \phi
   ~~
   +
    \imath
    \sin^2 \frac{\phi}{2}
  \right]
.
\label{eqn-standard-gamma2}
\end{align}

%
\label{S-standard-wave}

This standard choice of $k \equiv k_s$ leads 
 to a second order wave equation of the form
~
\begin{align}
  \partial_z^2 \vec{E} 
 +
  k_s^2 \vec{E}
 ~~
 +
  \imath
  k_n^2
  \gamma_s^2 
  \vec{E}
&=
  0
.
\label{eqn-standard-waveeqn-gamma2}
\end{align}
When factorized, 
 as briefly described in section \ref{S-waveequation}, 
 we get a pair of coupled,
 counter-propagating,
 first order equations.
These are
~
\begin{align}
  \partial_z E_{\pm}
&=
 \pm
  \imath k_s E_{\pm}
 \mp
  \frac{k_n^2}{2 k_s}
  \gamma_s^2
  \left(
    E_+ + E_-
  \right)
.
\label{eqn-standard-factored}
\end{align}

%
\label{S-standard-discussion}

Since the residual component $\imath \gamma_s^2$ on the RHS 
 of eqn. (\ref{eqn-standard-factored})
 contains a real part as well as an imaginary part,
 it is not pure loss.
The real part will impose oscillations
 on the field as it propagates, 
 thus altering the wave vector away from the 
 assumed value $k_s$.
However, 
 the real part is quadratic in $\phi$, 
 being $\propto \sin^2 \frac{\phi}{2}$,
 so for small losses the correction
 to the underlying propagation will be small.
If we rewrite eqn. (\ref{eqn-standard-factored})
 to incorporate the correction into the leading term, 
 we get
~
\begin{align}
  \partial_z E_{\pm}
&=
 \pm
  \imath 
  k_s
  \left[
    1 
   -
    \frac{k_n^2}{2 k_s^2}
    \sin^2 \frac{\phi}{2}
  \right]
  E_{\pm}
\qquad
 \mp
    \imath
  \frac{k_n^2}{2 k_s}
    \sin^2 \frac{\phi}{2}
  ~
  E_-
\nonumber
\\
& \qquad
 \mp
  \frac{1}{2}
  \frac{k_n^2}{2 k_s}
  \left[
    \sin \phi
  \right]
  \left(
    E_+ + E_-
  \right)
.
\label{eqn-standard-factored-v2}
\end{align}
As before, 
 the first term on the RHS is gives plane-wave-like propagation, 
 but now with a wave vector that differs from $k_s$.

I will now express the effective propagation wave vector
 in terms of $k_n$ and $\phi$.
To simplify the description, 
 I apply the 
 usually excellent \cite{Kinsler-2007josab} approximation
 that the effect of $E_-$ on the propagation can be ignored
 (i.e. set $E_-=0$).
Hence,
~
\begin{align}
  \partial_z E_{+}
&=
 +
  \imath 
  k_s'
  E_{+}
 -
  \frac{1}{2}
  \frac{k_n^2}{2 k_s}
    \sin \phi
   ~
    E_+
,
\\
\textrm{with} \qquad
  k'_s
&=
  k_n 
  \cos \frac{\phi}{2}
  \left[
    1 
   -
    \frac{1}{2}
    \tan^2 \frac{\phi}{2}
  \right]
.
\label{eqn-standard-factored-ksp}
\end{align}

For $\phi \ll 1$, 
 we then find that
~
\begin{align}
  {k'_s}^2
&\simeq
  k_n^2
    \cos \phi
.
\label{eqn-standard-factored-ksp2-approx}
\end{align}

Thus although I began with the standard definition,
 which assumes that the (forward-like) field
 will propagate with a wave vector $k \equiv k_s$, 
 we see instead that it propagates with 
 a wave vector $k \simeq k_n \sqrt{ \cos \phi }$.
As we will see,
 this approximation to the effective propagation wave vector
 is usually close to that of the alternative form
 discussed below; 
 the difference (for small loss) is of order $\phi^4$.

%
\label{S-standard-phase}

The standard phase velocity $v_p$ is
~
\begin{align}
  v_p^2
&=
  \frac{\omega^2}{k_s^2}
\qquad
=
  \frac{c^2}{n_0^2 \cos^2 \frac{\phi}{2}}
.
\label{eqn-standard-vp}
\end{align}

However, 
 if we were to use the effective propagation wave vector $k_s'$
 we would get a different answer; 
 in the case of the approximate form of 
 eqn. (\ref{eqn-standard-factored-ksp2-approx}), 
 it turns out the same as the alternate form given in the next section.

%
\label{S-standard-group}

The standard group velocity $v_g$ can be derived using
~
\begin{align}
  2 k_s \partial_\omega k_s
&=
  k_s^2
  \left[
    \frac{2}{n_0}
    \left( \partial_\omega n_0 \right)
   -
    \left( \partial_\omega \phi \right)
    \tan \frac{\phi}{2}
   +
    \frac{2}{\omega}
  \right]
.
\label{eqn-standard-2kdk}
\end{align}

Hence
~
\begin{align}
  v_g^{-1}
\qquad
=
  \partial_\omega k_s
&=
  \frac{k_s}
       {\omega} 
  \left[
    1
   +
    \frac{\omega}{n_0}
    \left( \partial_\omega n_0 \right)
   -
    \frac{\omega}{2}
    \left( \partial_\omega \phi \right)
    \tan \frac{\phi}{2}
  \right]
.
\label{eqn-standard-vg}
\end{align}

Just as for phase velocity,
 if we were to use the effective propagation wave vector $k_s'$, 
 we would get a different answer; 
 in the case of the approximate form of 
 eqn. (\ref{eqn-standard-factored-ksp2-approx}), 
 it turns out the same as the alternate form given 
 in the next section.

%
\section{The alternative form}\label{S-causal}

The alternative form for the wave vector, 
 based on the product $\epsilon \mu$,
 (i.e. the square of the refractive index,
 see eqn. (\ref{eqn-defs-causal-n})),
 is
~
\begin{align}
  k_c^2
&=
  \frac{\omega^2}{c^2}
  {
    \RealPart
      \left(
        n^2 
      \right)
  }
\qquad
=
  k_n^2
  \cos \phi 
\\
  k_c
&=
  k_n
  \sqrt{  \cos \phi }
.
\label{eqn-causal-k}
\end{align}
Thus $k_c$ is either real-valued or is pure imaginary.
Real values of $k_c$ correspond to a regime of propagating waves, 
 imaginary values to that of evanescent waves.
The phase velocity is then $u_p = c / n_c$, 
 and the (inverse) group velocity simply $u_g^{-1}=\frac{dk_c}{d\omega}$; 
 both will differ from the standard $v_p, v_g$, 
 and are given below.
Note that $k_c^2$ is related to $k_s^2$ by
~
\begin{align}
  \frac{k_c^2}{k_s^2}
&=
  \frac{k_n^2 \cos \phi}
       {k_n^2 \cos^2 \frac{\phi}{2}}
\qquad
= 
  1 - \tan^2 \frac{\phi}{2}
.
\end{align}

With this alternative choice, 
 it is simple to express $\beta^2$
 in terms of our wave vector $k_c^2$, 
~
\begin{align}
  \beta^2
&=
  k_c^2
 +
  k_n^2 
  \gamma_c^2  
,
\label{eqn-causal-beta2gamma2}
\end{align}
 with the residual behaviour described by 
~
\begin{align}
  \imath
  \gamma_c^2
&=
  \imath 
  \sin \phi
\label{eqn-causal-gamma2}
\qquad
=
  \imath
  \gamma_s^2
 +
  \sin^2 \frac{\phi}{2}
.
\end{align}
For small $\phi \ll 1$, 
 $\gamma_s$ and $\gamma_c$ 
 differ only by terms of order $\phi^2$.
Note that the loss-like part of the residual component
 (i.e. of $\ImagPart (\gamma_s^2)$ or $\ImagPart (\gamma_c^2)$)
 is the same for either form; 
 \emph{but that only this alternative form of $k$ (i.e. $k_c$)
 ensures that the residual component is purely lossy, 
 and will not change the spatial oscillations of the field 
 away from those of the propagation wave vector.}
However, 
 the alternative form of $k$ leads to the underlying propagation
 becoming evanescent if $\RealPart (n^2) < 0$.

%
\label{S-causal-wave}

With this choice of wave vector (i.e. $k \equiv k_c$), 
 the second order wave equation can be written
~
\begin{align}
  \partial_z^2 \vec{E} 
 +
  k_c^2 \vec{E}
 ~~
 +
  \imath
  k_n^2
  \gamma_c^2
  \vec{E}
&=
  0
\label{eqn-causal-waveeqn-gamma2}
\end{align}

When factorized, 
 as briefly described in section \ref{S-waveequation}, 
 we get 
~
\begin{align}
  \partial_z E_{\pm}
&=
 \pm
  \imath k_c E_{\pm}
 \mp
  \frac{k_n^2}{2 k_c}
  \gamma_c^2
    \sin \phi
  \left(
    E_+ + E_-
  \right)
.
\label{eqn-causal-factored}
\end{align}

%
\label{S-causal-phase}

The phase velocity $u_p$ is now
 \emph{faster} than for the standard definition, 
 being
~
\begin{align}
  u_p^2
&=
  \frac{\omega^2}{k_c^2}
\qquad
=
  v_p^2
  \left[
    1
   -
    \tan^2 \frac{\phi}{2}
  \right]^{-1}
.
\label{eqn-causal-vp-cf}
\end{align}

%
\label{S-causal-group}

The corresponding group velocity $u_g$ can be derived using
~
\begin{align}
  2 k_c \partial_\omega k_c
&=
  k_c^2
  \left[
    \frac{2}{n_0}
    \left( \partial_\omega n_0 \right)
   -
    \left( \partial_\omega \phi \right)
    \tan \phi
   +
    \frac{2}{\omega}
  \right]
.
\label{eqn-causal-2kdk}
\end{align}

Hence
~
\begin{align}
  u_g^{-1}
\qquad
=
  \partial_\omega k_c
&=
  \frac{k_c}{\omega}
  \left[
    1
   +
    \frac{\omega}{n_0}
    \left( \partial_\omega n_0 \right)
   -
    \frac{\omega}{2}
    \left( \partial_\omega \phi \right)
    \tan \phi
  \right]
.
\label{eqn-causal-vg}
\end{align}

Here the comparison of $u_g$ with the standard form $v_g$
 is less simple than for phase velocities: 
 the prefactors differ are different
 ($\sqrt{\cos(\phi)}$ compared to $\cos\frac{\phi}{2}$); 
 also the bracketed terms differ slightly
 (with $\tan \phi$ not $\tan\frac{\phi}{2}$).
However, 
 for $\phi<\pi/2$, 
 $\sqrt{\cos(\phi)} < \cos\frac{\phi}{2}$), 
 so that the group velocity $u_g$
 is faster than the standard $v_g$.

%
\section{Discussion}\label{S-comparisons}

As already noted, 
 for small losses the standard and alternative definitions of $n$ 
 (and also those of $k$)
 nearly coincide, 
 but they diverge as the loss increases.
Indeed, 
 for (e.g. strongly resonant) situations where  $\RealPart(n^2) <0$, 
 the underlying propagation
 (i.e. that defined by $k_s$ or $k_c$)
 can be of a completely different character.

%

The simplest case is the trivial one where where $\ImagPart(n^2)=0$.
Here $k_s^2 = k_c^2$, 
 and both are always positive; 
 both $\gamma_s^2$ and $\gamma_c^2$ are zero.
The descriptions are identical.

Next we add a small imaginary part to $n^2$,
 with $|\phi| \ll 1$, 
 so that $k_s$ and $k_c$ no longer match.
The loss-like part of the residual component
 is (as always) the same in both cases, 
 but a standard ($k_s$) description will be modified
 by an additional oscillation, 
 giving an effective wave vector comparable to $k_c$.
This is perhaps the most typical regime for device operation; 
 being either the low loss case of normal
 (positive phase velocity) propagation, 
 or the low loss case of NPV propagation.

\begin{figure}
\includegraphics[angle=-90,width=0.80\columnwidth]{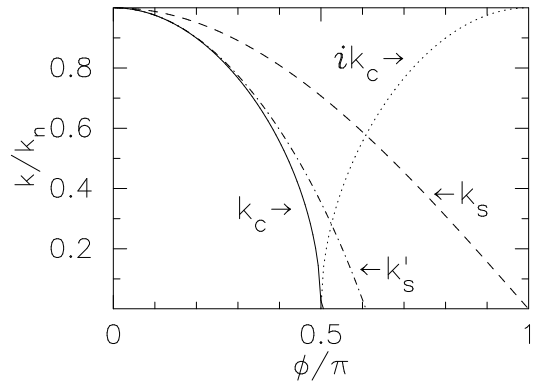}
\caption{
Comparison of $k$ values, 
 as a function of $\phi = \ArgPart(\epsilon\mu)$.
The causal choice $k_c$ is shown using a solid line
 when it is real valued, 
 and dotted when imaginary (``$\imath k_c$'');
 the standard choice ($k_s$) is given by the dashed line, 
 with the approximate corrected form ($k'_s$)  
 from eqn. (\ref{eqn-standard-factored-ksp2-approx})
 shown dot-dashed.
} \label{fig-comparek}
\end{figure}

As $\phi$ increases, 
 the two descriptions diverge, 
 as summarized on fig. \ref{fig-comparek}.
We see that 
 the standard description ($k \equiv k_s$) 
 gives qualitatively similar behaviour
 for all $|\phi| \le \pi$; 
 being one of a wave vector $k_s$ with added loss
 and a correction to achieve the true propagation wave vector.
Obviously, 
 the larger the $\phi$, 
 the larger the wave vector correction.

The alternative choice of $k \equiv k_c$ behaves differently.
When $|\phi|=\pi/2$, 
 i.e. when $n^2 = \ImagPart(n^2)$, 
 the wave vector $k_c$ vanishes, 
 giving no underlying oscillatory evolution as the field propagates.
The only evolution is that given by the residual component,
 i.e. the loss specified by $\ImagPart(n^2)$.
Then, 
 as $|\phi|$ increases further, 
 so that $\RealPart (n^2) = \RealPart (c^2 \epsilon \mu) < 0$,
 we find that $k_c$ takes on an imaginary value:
 this is just the case of plasmons, 
 where $\RealPart (\epsilon) \in \left(-\infty, 0 \right]$, 
 but $\RealPart (\mu) \in \left[0,\infty \right)$. 
Here 
 the imaginary $k_c$ means that underlying propagation becomes evanescent; 
 and any loss then acts in addition to that.

%
\label{S-causal-discussion}

Note that
 the loss in the alternative description 
 is simply $\ImagPart(n^2)$ --
 it differs from that used in the standard picture.
In particular note that this is \emph{not} identical 
 to the sum of the permittivity-based ``loss'' (i.e. $\ImagPart(\epsilon)$)
 and the permeability-based ``loss'' (i.e. $\ImagPart(\mu)$).
Further, 
 at least in the case of doubly passive media \cite{Kinsler-M-2008motl}, 
 $\ImagPart(n^2)<0$ is in fact a criterion for NPV; 
 i.e. \emph{loss} is a criterion for NPV.
More general statements on this relationship
 have been made when placing causality-based constraints
 on negative refractive index media
 using the Kramers-Kronig relations
 \cite{Stockman-2007prl,Kinsler-M-2008prl}.

%

Lastly,
 whichever choice of $k$ or $n$ we make,
 it depends only on the \emph{sum} of the complex phases
 of $\epsilon$ and $\mu$.
In contrast, 
 the summary given by \cite{Kinsler-M-2008motl} shows that 
 the NPV criteria of \cite{Depine-L-2004motl} also depends
 on the \emph{difference} of those phases.
This sensitivity arises
 because the presence of NPV depends on the relative 
 phases of the electric and magnetic fields; 
 however the second order wave equation 
 does not distinguish between the electric and magnetic responses, 
 considering only their nett effect on the selected field
 (here, the electric field $E$).

%
\section{Conclusion}\label{S-conclusion}

Here I have shown that the standard definition 
 for a real valued refractive index 
 (i.e. $n \equiv n_s = \RealPart ( \sqrt {n^2} )$)
 is only an approximation to the true real valued refractive index
 seen by a propagating optical pulse.
Instead, 
 the true propagation wave vector is based on 
 the alternate definition
 $n \equiv n_c = \sqrt{ \RealPart (n^2)}$.
This conclusion was reached by examining how fields
 are actually propagated by the widely used 
 electromagnetic second order wave equation, 
 in the case where when loss (or gain) is 
 treated as a modification to an underlying propagation
 based on a real-valued refractive index or wave vector.
Treatments of pulse propagation that use this alternative $n_c$
 (and hence $k_c$) 
 will not only be using wave vector that exactly matches the propagation, 
 but adjustments to that propagation will involve only gain or loss.
In contrast, 
 for the standard treatment based on $n_s, k_s$
 corrections to the spatial oscillation of the fields must be applied
 along with those for gain or loss.

%
\acknowledgments

I would like to thank M. W. McCall for useful discussions, 
 and G. H. C. New for helpful comments.
I also acknowledge financial support from the
 Engineering and Physical Sciences Research Council
 (EP/E031463/1).



\end{document}